\documentclass[epj]{svjour}
\usepackage{graphics}
\begin{document}
\title{Parity-Violation With Electrons: Theoretical Perspectives}
\author{M.J. Ramsey-Musolf\inst{1} % etc
\thanks{mjrm@caltech.edu}%
}                     % Do not remove
\institute{Kellogg Radiation Laboratory, California Institute of Technology, Pasadena, CA 91125 USA}
\date{Received: date / Revised version: date}
\abstract{
I review recent progress and developments in parity-violating electron scattering as it bears on three topics: strange quarks and hadron structure, electroweak radiative corrections, and physics beyond the Standard Model. I also discuss related developments in parity-conserving scattering with transversely polarized electrons as a probe of two-photon processes. I conclude with  a perspective on the future of the field. 
 % end of PACS codes
} %end of abstract
\maketitle
\section{Introduction}
\label{intro}
Parity-violating electron scattering (PVES) was once considered something of a specialized -- almost exotic -- subfield of nuclear physics. In the past decade, however, the field has made substantial advances and has become something of a mainstream area of research. This maturation of the field has been nicely summarized in the two PAVI meetings:  Mainz (2002) and  Grenoble (2004). My own involvement has now spanned a decade and a half, so I feel somewhat justified in providing a theoretical perspective on the status of the field. A rather extensive review can be found in my contribution to the Mainz PAVI proceedings\cite{Ramsey-Musolf:2003dd}, and this one will be somewhat abbreviated. Here, I will try to highlight what I think we have learned since the early 1990's and what I think may be the important directions for the future.

In reviewing the evolution of PVES, one can identify three rough eras: (1) the 1970's and 1980's, which  Wieman and Masterson have called \lq\lq  ancient history"\cite{Masterson:1996qi}; (2) the 1990's and first half of this decade, the \lq\lq modern era"; and (3) the future, lasting perhaps into the 2020's. The focus in the early era was on testing the neutral current structure of the Standard Model (SM). Pioneering experiments were carried out in both atomic PV (see Ref. \cite{Masterson:1996qi} for a review) as well as in the SLAC deep inelastic scattering (DIS) experiment\cite{Prescott:1978tm,Prescott:1979dh} that were followed by the Mainz $^8$Be quasielastic\cite{Heil:1989dz} and MIT-Bates $^{12}$C elastic\cite{Souder:1990ia} PVES measurements. In the 1990's, the structure of the SM neutral currents had been well established, so the emphasis shifted to using the SM neutral current as a probe of nucleon and nuclear structure. Here, the basic idea was that the weak neutral current depends on a different linear combination of the light quark currents than enters the electromagnetic current. Thus, by making a judicious choice of targets and kinematics, one can use PVES -- in combination with ordinary electron scattering -- to perform a flavor decomposition of the nucleon's vector current response. More recently, there has been a resurgence of interest in exploiting PVES to study the electroweak interaction itself. In this regard, the pioneering experiment has been the SLAC M\o ller experiment\cite{Anthony:2003ub} that will be followed by the equally demanding JLab Q-Weak experiment\cite{Armstrong:2003gp}. Looking down the road, one may see follow-ups to these measurements, possibly including  more precise versions of the original SLAC deep inelastic experiment at JLab or the M\o ller experiment at either JLab or the linear collider.

With this context in mind, let me try to summarize the recent developments in three fields, focusing primarily on the \lq\lq modern era": strange quarks and nucleon structure; radiative corrections, including both electroweak and QED; and the search for new physics.

\section{Strange Quarks}

The heyday of PVES has been dominated by strange quarks in the nucleon. Coming amidst the \lq\lq spin crisis" of the early 1990's, wherein there were indications that strange quarks carried a more substantial fraction of the nucleon's spin (and mass) than one might na\" ively think (for a review, see Ref.~\cite{Filippone:2001ux}), the question arose as to whether strange quarks might also play a substantial role in the nucleon's electromagnetic structure. Three developments -- two theoretical and one experimental -- catalyzed the use of PVES to address this question. Theoretically, Kaplan and Manohar observed how the weak neutral current (WNC) of the SM -- in conjunction with the electromagnetic (EM) current -- provided the tool needed to carry out the flavor decomposition of the nucleon's vector current structure\cite{Kaplan:1988ku}. Subsequently, Jaffe noted that dispersion theory analyses of the nucleon's isoscalar form factors implied large couplings of the nucleon to the $\phi(1020)$ and that this OZI-violation would imply sizeable strange quark vector current form factors as a consequence\cite{Jaffe:1989mj}. On the experimental side, McKeown showed that by measuring the backward angle PV asymmetry under conditions that were experimentally feasible, one would be sensitive to strange magnetic form factors of the magnitude implied by Jaffe's analysis\cite{Mckeown:1989ir}. (For other important, early work on this topic, see Refs.~\cite{Feinberg:1975cg,Walecka:1977us,Beck:1989tg,Napolitano:1990hi}).

The field was shortly off and running, resulting in the SAMPLE program at MIT-Bates\cite{Beise:2005qa,Spayde:2003nr,Ito:2003mr,Hasty:2001ep,Spayde:1999qg,Mueller:1997mt}, the HAPPEX\cite{Aniol:2004hp,Aniol:2000at} and G0\cite{Batigne:2004ma} experiments at JLab, and the A4 program at Mainz\cite{Maas:2004dh,Maas:2004ta,Baunack:2003ir}. A related development that I will not focus on here is the development of the JLab experiment designed to study the neutron distribution of $^{208}$Pb (see Ref.~\cite{Horowitz:1999fk} and references therein). The idea -- for which Donnelly, Dubach, and Sick  deserve the credit\cite{Donnelly:1989qs} -- is again to exploit the complementarity of the WNC and EM charge operators: the EM charge operator is sensitive primarily to protons, while the WNC primarily sees neutrons. The  details of each of these 
experiments has been discussed elsewhere in this meeting, so I will not comment on any further on the specifics.

While the G0, HAPPEX, and A4 experiments are not yet completed, the results obtained to date have taught us that the role of strange quarks in the nucleon vector current form factors is likely to be relatively less important than in the case of the nucleon mass and spin. In particular, the benchmark Jaffe predictions appear to be too large compared to experiment. At least, it is not consistent to have both sizable strange magnetism and strange electricity in the nucleon. The completion of the experimental program will definitively tell us whether there is room for either one to be relatively important. 

Theoretically, we have learned that strange quark dynamics are more subtle than the simplest pictures might suggest and that -- because the effects are not so large -- obtaining a quantitatively reliable description and realistic physical picture remains a challenging task. In terms of the simplest physical pictures that one might use to guide intuition, we have learned that neither vector meson dominance nor kaon cloud dominance is right. Vector meson dominance works well in describing the dynamics of pseudoscalar mesons; it has long been known, for example,  that the {\em a priori} unknown low energy constants (LECs) in the ${\cal O}(p^4)$ chiral Lagrangian for the octet of pseudoscalar mesons are well-described by vector meson dominance\cite{Ecker:1988te}. Although a similar {\em ansatz} allows one to obtain a reasonable fit to the nucleon isoscalar vector form factor in dispersion theory, Jaffe's generalization of it to the strange quark sector is now ruled out by experiment (for an up-date of the analysis in Ref.~\cite{Jaffe:1989mj}). 

The other picture that used to guide one's intuition is that of a kaon cloud around the nucleon. Its field theoretic description involves fluctuations of the nucleon into kaon-hyperon intermediate states whose effects are computed using loop graphs in perturbation theory. Now, one might rightfully object that given the large [${\cal O}(1)$] meson nucleon coupling constants ({\em e.g.}, 
$g_A\approx 1.26$), one has no basis for believing perturbation theory. Nevertheless, many theorists (myself included) proceeded to carry out such calculations anyway, using various models for the hadronic vertex form factors (for a list of references, see Ref.~\cite{Spayde:2003nr}). The motivation behind this mini-industry was an old computation of the nucleon's isovector form factors by Bethe and DeHoffman using pion loops\cite{bethe1955}. Despite the presence of the ${\cal O}(1)$ couplings, it worked, and so, based purely on this phenomenological success, people proceeded with the same approach for the strange form factors.

What most people did not realize, however, is that in the late 1950's, Federbush, Goldberger, and Treiman (FGT) showed -- using dispersion theory -- that the success of the Bethe and DeHoffman pion loop calculation was a big accident\cite{federbush}. In fact, the pion loop calculation as na\" ively performed does not even respect unitarity. When the latter requirement is imposed, the pion loop calculation falls short of the nucleon's isovector magnetic moment by half. FGT showed how to compute the pion cloud contribution correctly in dispersion theory by using measured strong interaction $\pi N$ scattering amplitudes and $\pi$ form factor data, effectively summing the loop graphs to all orders in the strong couplings. Remarkably, they obtained stunning agreement with the measured nucleon form factors. 

In the 1990's, my collaborators and I applied the FGT dispersion theory approach to the strange quark form factors, focusing on the kaon cloud contribution that under-girded so many model computations\cite{Hammer:1999uf,Hammer:1998rz,Ramsey-Musolf:1997qx,Musolf:1996qt}. Using experimentally obtained  $KN$ scattering amplitudes and $e^+ e^-$ data, we performed the all-orders summation of the kaon loop graphs analogous to the one carried out by FGT. We found that the resulting spectral functions for the strangeness vector form factors have a peak at invariant mass $t\approx m_\phi^2$. In short, the kaon cloud contribution is dominated by the $\phi(1020)$ resonance -- consistent with the earlier pole-dominance {\em ansatz}. 

This result has two implications. First, the plethora of kaon cloud model calculations are all wrong; only an all-orders summation is credible. Second, the kaon cloud cannot be the whole story, because the large effect implied by the $\phi$-resonance is ruled out by experiment. In principle, this situation is not incompatible with dispersion theory, which implies contributions from a tower of intermediate states. In practice, however, we have reached the end of the road, because there simply is not enough high-quality $e^+ e^-$ and strong interaction scattering data involving these other states\footnote{For a quark model treatment of part of this tower of states, see Ref.~\cite{Geiger:1996re}} . In the case of the isovector form factors, FGT got lucky because the pion cloud contribution -- when computed correctly -- saturated the experimental form factor result. There was no need to analyze other, higher mass intermediate states. The strange quark, in contrast, is more elusive. The PVES experiments suggest that contributions from other intermediate states cancel against the kaon cloud, but theoretically, we simply cannot compute these canceling effects. Parenthetically, my experience with this dispersion theory analysis leads me to view simple meson cloud model computations of other nucleon structure effects with a degree of skepticism.

It is reasonable to ask, however, why one has to rely on dispersion theory in the \lq\lq modern era". Indeed, the effects of pseudoscalar mesons in various aspects of nucleon structure -- such as the isovector form factors or polarizablities -- have been treated with considerable success in chiral perturbation theory (ChPT). In ChPT, any observable is given by the sum of a loop contribution and an LEC, or counterterm\footnote{Usually, one takes for the loop contribution only the part of a loop result that is non-analytic in external momenta or masses that cannot be written down in a Lagrangian.}. Even though one never carries out an all-orders loop calculation as dispersion theory suggests one should, this approach is consistent with dispersion theory because of the presence of the LEC contribution and because for low-energy processes, higher-order effects are suppressed by $p/\Lambda$, where $\Lambda$ is either the nucleon mass or chiral symmetry-breaking scale. The actual value of the LEC, however, cannot be predicted theoretically; it has to be taken from experiment. To the extent that one has enough independent experiments to determine all the LECs at a given order in $p/\Lambda$, one can then make predictions for other observables. 

The problem for strange quarks is that we don't have the required set of independent experiments. As Ito and I pointed out some time ago, the leading-order LEC's contain an SU(3)-singlet component that cannot be taken from any existent experiments\cite{Musolf:1996zv}. In order to determine these constants, one has to measure the very strange quark form factors that one would like to predict -- a situation of circular logic. Formally, an exception occurs in the case of the strange magnetic radius, $\langle r_s^2\rangle_M$. As shown by Meissner, Hemmert, and Steininger, the leading-order contribution is entirely non-analytic, so that a counterterm-free prediction can be obtained from a one-loop computation\cite{Hemmert:1998pi}. Unfortunately, the chiral expansion for SU(3) is slowly-converging, since the relevant expansion parameter is $m_k/\Lambda\sim 1/2$. Thus, one might worry that higher-order effects are not negligible. In fact, my collaborators and I showed that -- in the case of the strange magnetic radius -- the next-to-leading order (NLO) loop contribution cancels most of the LO loop effect, exposing one to a dependence on the NLO LEC, $b_s^r$\cite{Hammer:2002ei}. The resulting expression for $\langle r_s^2\rangle_M$ is
\begin{equation}
\label{eq:magrad}
\langle r_s^2\rangle_M = -\left[0.04+0.3\ b_s^r\right]\ {\rm fm}^2\ \ \ ,
\end{equation}
where the first term on the RHS gives the loop contribution evaluated at a renormalization scale $\mu=1$ GeV and the second term is the NLO counterterm contribution. Naturalness considerations suggest that $b_s^r$ should have magnitude of order unity, implying that the second term on the RHS of Eq.~(\ref{eq:magrad}) dominates over the first. Since the precise value of $b_s^r$ cannot be determined except by measuring the strange magnetic radius itself, one is back to the original problem. It seems there is no free lunch with strange quarks. The situation is not entirely bleak, as one can get some indications of the size of the LEC from either dispersion theory or lattice QCD, and a reasonable (though not rigorous) range of predictions can be obtained in this case. Indeed, the final SAMPLE result for the strange magnetic moment -- which requires extrapolation to $Q^2=0$ by using the magnetic radius -- includes this range in its quoted error\cite{Spayde:2003nr}.

It goes without saying that one would like to carry out reliable, microscopic calculations of the strangeness form factors in QCD, and the lattice is our only tool for doing so. Some time ago, the Kentucky-Adelaide collaboration obtained a non-zero signal for the form factors in a quenched calculation, the results of which suggested a negative strangeness magnetic moment\cite{Dong:1997xr}.
More recently, the authors of Ref.~\cite{Lewis:2002ix} performed an quenched computation and found no evidence for non-zero strangeness form factors. The two computations differed in the number of gauge configurations and $Z_2$ noise vectors employed. To my knowledge, the two groups have not yet sorted out the reasons for the difference in their two results. As we also heard in this meeting, D. Leinweber and collaborators have exploited quenched lattice results and the assumption of charge symmetry to predict the strangeness magnetic moment. The details may be found in his talk, but the approach has successfully reproduced the measured octet baryon magnetic moments. The predicted strangeness magnetic moment is negative: $\mu_s=-0.051\pm 0.021$\cite{Leinweber:2004tc}. The conundrum for this prediction as well as for the previously reported lattice results is that the PVES experiments suggest a {\em positive} value for $\mu_s$ (see, {\em e.g.}, Ref.~\cite{Maas:2004dh}). If these indications are solidified with the final results from the experiments, more work will be needed to understand the sign of $\mu_s$ from first principles in QCD (for an observations regarding the possible role of $C$ transformation properties, see Ref.~\cite{Ji:1995rd}) .

Finally, let me note that hadron models may provide some insights into the strangeness form factors. At present, one model -- the chiral quark soliton model -- gives the only prediction for a positive $\mu_s$ that I am aware of\cite{Silva:2002ej}. In light of all other theoretical work that seems to suggest a negative sign, this prediction must be taken seriously. Since I am not an expert on this model,  I cannot comment in detail on what it may mean for the QCD dynamics of strange quarks. It is suggestive, however, that the topology of the QCD vacuum has a nontrivial impact on sea quark dynamics.

\section{Radiative Corrections}

A significant consideration in the theoretical interpretation of the PV asymmetries and their implications for strange quarks has been contributions from electroweak radiative corrections. In high energy processes, one can compute these corrections with a high degree of reliability. At the low energies relevant to the PVES experiments, the situation is quite different due to the interplay of the strong interaction and higher-order electroweak effects. The electroweak Ward Identities protect the weak neutral vector current from  strong interaction renormalization, but the the axial vector current can experience substantial effects. Since the axial vector response contributes to a generic PV asymmetry, one has to take into account the {\em uncertainties} associated with these axial vector radiative correction effects. Uncertainties also arise in the vector channel {\em via} box diagrams in which two electroweak gauge bosons are exchanged; no symmetry protects one from QCD effects in this case. Fortunately, however, the potentially largest uncertainties, which would appear in the $Z-\gamma$ box graphs, are fortuitously suppressed by $1-4\sin^2\theta_W\sim 0.1$\cite{Marciano:1983ss,Ramsey-Musolf:1999qk}. Consequently, one's primary concern involves the axial vector channel.

As part of my Ph.D. research -- carried out in collaboration with Barry Holstein -- I showed that one class of axial vector corrections -- the so-called \lq\lq anapole moment" terms -- can be both surprisingly large and theoretically uncertain\cite{Musolf:1990ts}. These corrections involve the exchange of a virtual $\gamma$ between the lepton and hadron target, so their effect in neutrino-hadron scattering is suppressed by an additional power of $\alpha/4\pi$ compared to the effect in PVES. Consequently, it is useful to distinguish between the axial vector form factor as measured in electron scattering, $G_A^e$, from the corresponding form factor probed by neutrinos, $G_A^\nu$. As Holstein and I pointed out in Ref.~\cite{Musolf:1990ts}, the uncertainties associated with the anapole contributions to $G_A^e$ are at least as large as the contribution expected from strange quarks (as extrapolated from polarized deep inelastic scattering), so PVES provides a rather poor probe of axial vector strangeness. Neutrino scattering is theoretically much cleaner, since the radiative correction uncertainties are insignificant. 

Subsequent to this work, Donnelly and I observed that the radiative correction uncertainties in $G_A^e$ could mimic the effect of strange magnetism on the elastic, ${\vec e}p$ PV asymmetry\cite{Musolf:1992xm}. Consequently, without a better handle on $G_A^e$, one faced an intrinsic, theoretical uncertainty in the value of $G_M^s$ that could be extracted from PVES experiments. More generally, this kind of delicate interplay between various theoretical inputs -- including radiative corrections -- and the extracted values of the strange quark form factors was laid out in Ref.~\cite{Musolf:1993tb}. Fortunately, Hadjimichael, Poulis, and Donnelly  observed that by measuring the PV asymmetry in quasieslastic (QE) ${\vec e}D$ scattering, one could obtain a different linear combination of $G_M^s$ and $G_A^e$ than enters elastic ${\vec e}p$ scattering, thereby facilitating an {\em experimental} separation of both\cite{Hadjmichael:1991be}. With this goal in mind, the SAMPLE Collaboration measured both asymmetries. As initially reported, the results implied a value for $G_A^e$ consistent with zero and a value of $G_M^s$ that was somewhat positive\cite{Hasty:2001ep}. The result for $G_A^e$ was particularly surprising. Originally, Holstein and I predicted that radiative corrections would reduce $G_A^e$ by roughly 30\%, but the SAMPLE results suggested a 100\% reduction. This surprise set off a flurry of theoretical efforts to explain the large effect. My collaborators and I up-dated the earlier work with Holstein using heavy baryon chiral perturbation theory and found little room for a larger effect than originally predicted\cite{Zhu:2000gn}. Others explored the $Q^2$-dependence of the anapole form factor\cite{Maekawa:2000qz,Maekawa:2000bd}, quark model estimates of the anapole contribution\cite{Riska:2000qw}, and nuclear PV effects in the ${\vec e}D$ reaction\cite{Schiavilla:2002uc,Liu:2002bq}. Despite this hard work, no one could explain the SAMPLE result. One remaining suspect remained to be analyzed: the $Z-\gamma$ box contribution. In contrast to its effect in the vector channel, its contribution to the axial response is not  $1-4\sin^2\theta_W$ suppressed, so it was thought that a large effect might be present here. Subsequently, however, the SAMPLE collaboration reanalyzed the deuterium data and found three previously underestimated corrections\cite{Ito:2003mr}. The largest involved neutral pion backgrounds. The resulting shifts in the PV QE asymmetry was just enough to change the extracted value of $G_A^e$, bringing it into agreement with the original Musolf and Holstein prediction. 

What we have learned from this experience is that non-perturbative QCD effects on electroweak radiative corrections in the axial vector channel can be significant, and one must take them into account somehow. Doing so has become important not only for the strange quark program, but also for interpretation of the G0 measurement of the PV $N\to\Delta$ asymmetry. Here, the same kind of anapole effects that enter the elastic asymmetry can be as significant\cite{Zhu:2001re}. Consequently, the normalization of the axial vector transition form factor probed by PVES, $G_A^{\Delta,e}$ at $Q^2=0$, will contain a theoretical radiative correction uncertainty of order 10-20\%. As Shi-lin Zhu and I recently showed, a determination of these corrections could be achieved by combining an experimental value for $G_A^{\Delta,e}(Q^2=0)$ with the off-diagonal Goldberger-Treiman relation (ODGTR), as the latter predicts the value of $G_A^\Delta(Q^2=0)$ in the absence of these corrections\cite{Zhu:2002kh}:
\begin{equation}
\label{eq:odgtr}
G_A^\Delta(0) = \sqrt{\frac{2}{3}}\frac{g_{\pi N\Delta}F_\pi}{m_N}\left(1-\Delta_\pi\right)\ \ \ ,
\end{equation}
where $g_{\pi N\Delta}$ is the strong $\pi N\Delta$ coupling constant that is known with 5\% accuracy from $\pi N$ scattering in the resonance region, $F_\pi$ is the pion-decay constant, and $\Delta_\pi$ is a chiral correction that is of order a few percent. Since $G_A^{\Delta, e}(0)$ differs from $G_A^\Delta(0)$ by the large electroweak radiative corrections, a measurement of the former with PVES will primarily probe these radiative corrections.

In addition, a new effect arises in the inelastic asymmetry that does not occur for elastic scattering. Specifically, the asymmetry no longer vanishes at $Q^2=0$. This result is a consequence of Siegert's theorem\cite{Siegert:1937yt}, which implies that elastic matrix elements of the $E1$ operator vanish (even in the presence of parity-mixing in the initial and/or final states), but that they need not do so for inelastic reactions. The resulting, low-$Q^2$ $N\to\Delta$ asymmetry has the form
\begin{equation}
A_{LR}^{N\to\Delta}(Q^2=0) = -2 \frac{d_\Delta}{C_3^V}{m_N\over\Lambda_\chi}+\cdots
\end{equation}
where $C_3^V$ is the transition magnetic form factor, $d_\Delta$ parameterizes the $N\to\Delta$ $E1$ amplitude, and the \lq\lq $+\cdots$ indicate higher order chiral corrections.

This effect for the $N\to\Delta$ asymmetry was first pointed out by in Ref.~\cite{Zhu:2001br}, where we argued that measuring $d_\Delta$ could be important for two reasons. First, without knowing its value, any attempt to determine the $Q^2$-dependence of   $G_A^{\Delta,e}$ with PVES could be confused by this term. Second, $d_\Delta$ is the neutral current analog of the $E1$ amplitudes for PV, electromagnetic hyperon decays, such as ${\vec\Sigma}^+\to p\gamma$. The asymmetries associated with the latter are of order four times larger in magnitude than one would expect based on simple symmetry considerations -- a puzzle that has largely eluded explanation. Studying the neutral current analog ({\em i.e.}, $d_\Delta$) could, in principle, shed new light on this old problem.  New efforts are underway to develop experiments that would determine $d_\Delta$.

More recently, another aspect of radiative corrections -- those involving the exchange of two photons -- has grabbed the attention of the PVES community. While this effect is pure QED and conserves parity, one manifestation of it can be measured using similar methods to those used in the PV experiments. Specifically, by scattering transversely polarized electrons from a given target and measuring the transverse-spin asymmetry (or vector analyzing power), $A_n$, one probes the imaginary part of the two-photon exchange amplitude, ${\cal M}_{\gamma\gamma}$. The vector analyzing power (VAP) has been measured by the SAMPLE Collaboration at relatively low energy\cite{Wells:2000rx} and by the Mainz A4 Collaboration at higher energy\cite{Maas:2004pd}. The SAMPLE result differs substantially from an old, potential scattering prediction carried out by Mott\cite{mott}, leading to considerable theoretical interest in this observable. Although the VAP has been
discussed by several others in this meeting, I would like to comment on why I think it is an important topic to pursue experimentally and theoretically. 

By now it is well-known that a proper treatment of the real part of ${\cal M}_{\gamma\gamma}$ may help resolve the apparent differences in the $Q^2$-dependence of the proton electric form factor as determined  polarization transfer experiments {\em vs.} Rosenbluth separation (for a discussion, see, {\em e.g.}, Refs.~\cite{Arrington:2003qk,Guichon:2003qm} and references therein). Clearly, having an experimental test of theoretical calculations of ${\cal M}_{\gamma\gamma}$ is important from this standpoint, and the VAP provides such a test. It is also important for our understanding of electroweak radiative corrections. Indeed, the kinds of two-boson exchange box graphs mentioned above have to be computed theoretically in order to arrive at Standard Model predictions for various electroweak observables. In particular, the dominant theoretical uncertainty associated with the interpretation of neutron and nuclear $\beta$-decay involves such graphs, where one of the exchanged bosons is a $W^\pm$ and the other is a $\gamma$. In order to extract a value for the CKM matrix element $V_{ud}$ from these measurements, one must compute the $W\gamma$ box contribution. The theoretical machinery needed for this computation is the same as required for ${\cal M}_{\gamma\gamma}$, but there is no hope of ever measuring ${\cal M}_{W\gamma}$ directly. Hence, the VAP may provide our only experimental test of the theoretical framework used in computing this important electroweak radiative correction.

The $\beta$-decay correction applies to a very low-energy process, for which the use of effective field theory (EFT) might be particularly suitable. Similarly, the recent SAMPLE measurement of the VAP was performed in the low-energy/EFT domain. With this context in mind, my student L. Diaconescu and I recently computed the VAP for low-energy, elastic $ep$ scattering using an EFT approach. Our primary objective was to determine if one could use EFT to resolve the discrepancy between the SAMPLE result and the old potential scattering calculation performed by Mott. If successful, we would have some confidence in applying the same framework to the electroweak box correction. The details of our calculation can be found in Ref.~\cite{Diaconescu:2004aa}. Let me comment, however, on a few aspects of the calculation. 

First, given the technical challenge in carrying out analytic computations, we decided to work initially with an EFT involving only electrons, photons, and nucleons. In the heavy baryon formalism, the nucleon is essentially a \lq\lq static" source, and there is no large momentum associated with this degree of freedom. In this EFT, the pion is integrated out, being treated as \lq\lq heavy". While this assumption would certainly hold for the $\beta$-decay correction, it is admittedly questionable for the VAP at the SAMPLE kinematics. Nonetheless, we explored how much of the experimentally observed VAP could be accounted for by this simplest EFT. In this case, $A_n$ has a power series expansion in $p/M$, where $M$ is the nucleon mass and $p$ is either the incident electron energy or electron mass. The Mott computation corresponds to the ${\cal O}(p/M)^0$ contribution -- the one that survives for an infinitely heavy target. We showed that, working to ${\cal O}(p/M)^2$, one can make a parameter-free prediction for the VAP. To this order, there are no unknown LEC's, and the VAP arises entirely from a one-loop effect. The first unknown constants appear at ${\cal O}(p/M)^4$. To obtain a consistent computation to ${\cal O}(p/M)^2$, one must include both the nucleon magnetic moment and charge radius at the $\gamma NN$ vertices in the loop, as well as \lq\lq kinetic" terms in the nucleon propagator.

To our surprise, we found that the ${\cal O}(p/M)^2$ VAP agrees beautifully with the SAMPLE result. Given that the momentum transfer is of order $m_\pi$, one would have expected that inclusion of pions  would be necessary to produce agreement. What the result suggestions, however, is that for this particular observable, pions do not appear to play a significant role at these kinematics. As a corollary, it seems reasonable that using the same EFT (without dynamical pions) at the lower-energies relevant for $\beta$-decay will produce a reliable result for that process. Carrying out the latter computation is clearly a task for the future. On the other hand, taking the EFT expression for $A_n$ to the higher energies relevant to the Mainz VAP measurement leads to significant disagreement -- not a surprising outcome given that the Mainz energies are well beyond the limit of this EFT. For this domain, one must likely include dynamical pions and the $\Delta$-resonance. I don't know how to do that in a model-independent way that retains the systematic power counting of EFT, so some other approach may be necessary. B. Pasquini has reported on one such attempt using the MAID program\cite{Pasquini:2004pv}. As I understand the results, the MAID calculation comes closer to the Mainz results than the EFT calculation, but fails to reproduce the lower energy SAMPLE result that can be explained using EFT (for other model computations, see, {\em e.g.}, Refs.~\cite{Afanasev:2004pu,Afanasev:2004hp}). It is clearly of interest to understand the reasons behind these differing theoretical results, and having more experimental data in the low- to medium-energy domain would be quite helpful.

\section{New Physics}

A subject that is close to my heart -- but one that I will spend less time on here -- is the use of PVES to search for physics beyond the Standard Model. Two experiments are currently on the books: the E158 M\o ller experiment at SLAC\cite{Anthony:2003ub} that has finished its data taking and the Q-Weak experiment that will take place in the future at JLab\cite{Armstrong:2003gp}. The beauty of these two experiments is that they are highly precise, that they are being carried out at essentially the same $Q^2$, and that they are theoretically \lq\lq clean" probes of the Standard Model and its extensions. Indeed, these two experiments, when taken in tandem with the results of the Cesium atomic PV experiment, provide a unique diagnostic of new physics. My collaborators and I have recently illustrated this point in several papers, focusing particularly intently on supersymmetric extensions of the Standard Model\cite{Erler:2003yk,Kurylov:2003zh,Kurylov:2003xa,Ramsey-Musolf:2000qn,Ramsey-Musolf:1999qk}. I refer the reader to those papers for details. Even in the absence of new physics effects, the two experiments will provide the most precise determinations of the running $\sin^2\theta_W$ at a scale below the $Z^0$-pole (see, {\em e.g.}, Refs.~\cite{Czarnecki:1995fw,Erler:2003yk,Ferroglia:2003wa,Erler:2004in}). 

Looking to the future, the diagnostic power of the PVES experiments could be amplified with additional measurements. A more precise version of the M\o ller experiment is under consideration for JLab in its post-12 GeV upgrade phase. As I understand it, this experiment would be quite challenging, but several talented people are working on ways to meet this challenge. A second possibility would be to carry out a more precise version of the SLAC deep inelastic experiment. Recently, a proposal to do so was given high marks by the SLAC E-PAC, but the funds to carry out this experiment do not seem to be available. There is also considerable interest in performing a lower-energy version of the PV deep inelastic, or
\lq\lq DIS-Parity" experiment at the up-graded JLab. In this case, sorting out the theoretical implications of a precision measurement would be more involved than for the SLAC kinematics, primarily because of potential higher-twist (HT)  contributions. The SLAC $Q^2$ would be sufficiently large that one could neglect these HT effects and perform a precision electroweak test. On the other hand, \lq\lq twist pollution" will creep in at the JLab kinematics, and we do not have enough information on HT contributions to make definitive statements about their size at these kinematics. It may be that carrying out multiple measurements at different momentum transfers would allow one to simultaneously constrain the degree of twist pollution and deviations from the Standard Model electroweak contribution. Given that the figure of merit for the deep inelastic measurements is relatively large, such a program may be realistic.

Theoretically, several questions pertaining to the deep inelastic asymmetry remain to be studied extensively. In the case of HT, for example, one does not know what QCD predicts for the evolution of the corresponding structure functions. Similarly, the degree to which isospin violation in the leading twist parton distribution functions would affect the asymmetry is an interesting question. As we heard in K. McFarland's talk, the currently favored explanation of the NuTeV anomaly\cite{Zeller:2001hh} is just such isospin violation. To be a viable one, this violation would have to be large, so that one might expect similarly large effects on the deep inelastic asymmetry. Finally, there is the question as to what combination of measurements would yield the most useful information on the electroweak sector of the Standard Model as well as on the aforementioned aspects of nucleon structure. Clearly, there is considerable room for future theoretical work on this topic.

\section{Conclusions} 

It seems to me that the \lq\lq modern era" of PVES experiments will be coming to an end in the next few years. With the conclusion of the SAMPLE, HAPPEX, G0, and A4 programs, the experimental questions about the size of the strange quark form factors and axial vector radiative corrections will have been settled; the $^{208}$Pb measurement will have provided new information about the distribution of neutrons in heavy nuclei; and the first generation of precision electroweak tests (E158 and Q-Weak) will have produced results. I believe that those who have been involved in this field will be able to look back with considerable satisfaction at the unique and varied physics that this subfield of nuclear physics has illuminated. At the same time, there will be ample grist for the theoretical mill. If the present indications of a positive strangeness magnetic moment persist, it will be a challenge to provide a QCD explanation for this result and to understand the dynamics behind it. Similarly, if G0 and possible new experiments are able to cleanly separate the $G_A^{\Delta,\ e}$ and $d_\Delta$ contributions to the inelastic asymmetry, then new confrontations with hadron structure theory will be available. Finally, comparisons of first results from the LHC with information obtained from precision measurements -- such as Cesium atomic PV, E158, Q-Weak, and possible second generation progeny -- may help us determine which  of the currently popular extensions of the Standard Model -- if any -- are most viable. In short, it appears that we have hardly heard the last word from parity violation.

\section {Acknowledgments}
I am indebted to my many theoretical and experimental collaborators and colleagues who are too numerous to list and who, over the years, have helped me understand various aspects of this field with greater clarity and insight.
This work was supported in part by U.S. Department of Energy 
DE-FG03-02ER41215  and by a National Science
Foundation Grant PHY00-71856.

%\begin{figure}
%\resizebox{0.75\textwidth}{!}{%
%  \includegraphics{leer.eps}
%\vspace{5cm}       % Give the correct figure height in cm
%\end{figure}
%\begin{figure*}
%\vspace*{5cm}       % Give the correct figure height in cm
%\end{figure*}
%\begin{table}
%\begin{tabular}{lll}
%\hline\noalign{\smallskip}
%\end{tabular}
% Or use
%\vspace*{5cm}  % with the correct table height
%\end{table}
%
% BibTeX users please use
% \bibliographystyle{}
% \bibliography{}

\begin{thebibliography}{}
%\cite{Ramsey-Musolf:2003dd}
\bibitem{Ramsey-Musolf:2003dd}
M.~J.~Ramsey-Musolf,
%``Parity-Violating Electron Scattering: How Strange a Future,''
arXiv:nucl-th/0302049.
%%CITATION = NUCL-TH 0302049;%%

%\cite{Masterson:1996qi}
\bibitem{Masterson:1996qi}
B.~P.~Masterson and C.~E.~Wieman, in {\em Precision Tests of the Standard Model}, P. Langacker, Ed. (World Scientific, Singapore, 1995), Ch. IV.D.
%``Atomic parity nonconservation experiments,''
%\href{http://www.slac.stanford.edu/spires/find/hep/www?irn=3465217}{SPIRES entry}

%\cite{Prescott:1978tm}
\bibitem{Prescott:1978tm}
C.~Y.~Prescott {\it et al.},
%``Parity Non-Conservation In Inelastic Electron Scattering,''
Phys.\ Lett.\ B {\bf 77}, 347 (1978).
%%CITATION = PHLTA,B77,347;%%

%\cite{Prescott:1979dh}
\bibitem{Prescott:1979dh}
C.~Y.~Prescott {\it et al.},
%``Further Measurements Of Parity Nonconservation In Inelastic Electron
%Scattering,''
Phys.\ Lett.\ B {\bf 84}, 524 (1979).
%%CITATION = PHLTA,B84,524;%%

%\cite{Heil:1989dz}
\bibitem{Heil:1989dz}
W.~Heil {\it et al.},
%``Improved Limits On The Weak, Neutral, Hadronic Axial Vector Coupling
%Constants From Quasielastic Scattering Of Polarized Electrons,''
Nucl.\ Phys.\ B {\bf 327}, 1 (1989).
%%CITATION = NUPHA,B327,1;%%

%\cite{Souder:1990ia}
\bibitem{Souder:1990ia}
P.~A.~Souder {\it et al.},
%``Measurement Of Parity Violation In The Elastic Scattering Of Polarized
%Electrons From C-12,''
Phys.\ Rev.\ Lett.\  {\bf 65}, 694 (1990).
%%CITATION = PRLTA,65,694;%%

%\cite{Anthony:2003ub}
\bibitem{Anthony:2003ub}
P.~L.~Anthony {\it et al.}  [SLAC E158 Collaboration],
%``Observation of parity nonconservation in Moeller scattering,''
Phys.\ Rev.\ Lett.\  {\bf 92}, 181602 (2004)
[arXiv:hep-ex/0312035].
%%CITATION = HEP-EX 0312035;%%

%\cite{Armstrong:2003gp}
\bibitem{Armstrong:2003gp}
D.~S.~Armstrong {\it et al.}  [Qweak Collaboration],
%``Qweak: A precision measurement of the proton's weak charge,''
AIP Conf.\ Proc.\  {\bf 698}, 172 (2004)
[arXiv:hep-ex/0308049].
%%CITATION = HEP-EX 0308049;%%

%\cite{Filippone:2001ux}
\bibitem{Filippone:2001ux}
B.~W.~Filippone and X.~D.~Ji,
%``The spin structure of the nucleon,''
Adv.\ Nucl.\ Phys.\  {\bf 26}, 1 (2001)
[arXiv:hep-ph/0101224].
%%CITATION = HEP-PH 0101224;%%

%\cite{Kaplan:1988ku}
\bibitem{Kaplan:1988ku}
D.~B.~Kaplan and A.~Manohar,
%``Strange Matrix Elements In The Proton From Neutral Current Experiments,''
Nucl.\ Phys.\ B {\bf 310}, 527 (1988).
%%CITATION = NUPHA,B310,527;%%

%\cite{Jaffe:1989mj}
\bibitem{Jaffe:1989mj}
R.~L.~Jaffe,
%``Stranger Than Fiction: The Strangeness Radius And Magnetic Moment Of The
%Nucleon,''
Phys.\ Lett.\ B {\bf 229}, 275 (1989).
%%CITATION = PHLTA,B229,275;%%

%\cite{Mckeown:1989ir}
\bibitem{Mckeown:1989ir}
R.~D.~Mckeown,
%``Sensitivity Of Polarized Elastic Electron Proton Scattering To The Anomalous
%Baryon Number Magnetic Moment,''
Phys.\ Lett.\ B {\bf 219}, 140 (1989).
%%CITATION = PHLTA,B219,140;%%


%\cite{Feinberg:1975cg}
\bibitem{Feinberg:1975cg}
G.~Feinberg,
%``Polarized Electron - Nucleus Scattering And Parity Violating Neutral Current
%Interactions,''
Phys.\ Rev.\ D {\bf 12}, 3575 (1975)
[Erratum-ibid.\ D {\bf 13}, 2164 (1976)].
%%CITATION = PHRVA,D12,3575;%%

%\cite{Walecka:1977us}
\bibitem{Walecka:1977us}
J.~D.~Walecka,
%``Semileptonic Weak And Electromagnetic Interactions In Nuclei: Parity
%Violations In Electron Scattering And Weak Neutral Currents,''
Nucl.\ Phys.\ A {\bf 285}, 349 (1977).
%%CITATION = NUPHA,A285,349;%%

%\cite{Beck:1989tg}
\bibitem{Beck:1989tg}
D.~H.~Beck,
%``Strange Quark Vector Currents And Parity Violating Electron Scattering From
%The Nucleon And From Nuclei,''
Phys.\ Rev.\ D {\bf 39} (1989) 3248.
%%CITATION = PHRVA,D39,3248;%%

%\cite{Napolitano:1990hi}
\bibitem{Napolitano:1990hi}
J.~Napolitano,
%``Measuring The Strangeness Radius Of The Proton,''
Phys.\ Rev.\ C {\bf 43}, 1473 (1991).
%%CITATION = PHRVA,C43,1473;%%

%\cite{Beise:2005qa}
\bibitem{Beise:2005qa}
E.~J.~Beise, M.~L.~Pitt and D.~T.~Spayde,
%``The SAMPLE experiment and weak nucleon structure,''
Prog.\ Part.\ Nucl.\ Phys.\  {\bf 54}, 289 (2005).
%%CITATION = PPNPD,54,289;%%

%\cite{Spayde:2003nr}
\bibitem{Spayde:2003nr}
D.~T.~Spayde {\it et al.}  [SAMPLE Collaboration],
%``The strange quark contribution to the proton's magnetic moment,''
Phys.\ Lett.\ B {\bf 583}, 79 (2004)
[arXiv:nucl-ex/0312016].
%%CITATION = NUCL-EX 0312016;%%

%\cite{Ito:2003mr}
\bibitem{Ito:2003mr}
T.~M.~Ito {\it et al.}  [SAMPLE Collaboration],
%``Parity-violating electron deuteron scattering and the proton's neutral weak
%axial vector form factor,''
Phys.\ Rev.\ Lett.\  {\bf 92}, 102003 (2004)
[arXiv:nucl-ex/0310001].
%%CITATION = NUCL-EX 0310001;%%

%\cite{Hasty:2001ep}
\bibitem{Hasty:2001ep}
R.~Hasty {\it et al.}  [SAMPLE Collaboration],
%``Strange magnetism and the anapole structure of the proton,''
Science {\bf 290}, 2117 (2000)
[arXiv:nucl-ex/0102001].
%%CITATION = NUCL-EX 0102001;%%

%\cite{Spayde:1999qg}
\bibitem{Spayde:1999qg}
D.~T.~Spayde {\it et al.}  [SAMPLE Collaboration],
%``Parity violation in elastic electron proton scattering and the proton's
%strange magnetic form-factor,''
Phys.\ Rev.\ Lett.\  {\bf 84}, 1106 (2000)
[arXiv:nucl-ex/9909010].
%%CITATION = NUCL-EX 9909010;%%

%\cite{Mueller:1997mt}
\bibitem{Mueller:1997mt}
B.~Mueller {\it et al.}  [SAMPLE Collaboration],
%``Measurement of the proton's neutral weak magnetic form factor,''
Phys.\ Rev.\ Lett.\  {\bf 78}, 3824 (1997)
[arXiv:nucl-ex/9702004].
%%CITATION = NUCL-EX 9702004;%%
%\cite{Aniol:2004hp}
\bibitem{Aniol:2004hp}
K.~A.~Aniol {\it et al.}  [HAPPEX Collaboration],
%``Parity-violating electroweak asymmetry in e(pol.) p scattering,''
Phys.\ Rev.\ C {\bf 69}, 065501 (2004)
[arXiv:nucl-ex/0402004].
%%CITATION = NUCL-EX 0402004;%%

%\cite{Aniol:2000at}
\bibitem{Aniol:2000at}
K.~A.~Aniol {\it et al.}  [HAPPEX Collaboration],
%``New measurement of parity violation in elastic electron proton  scattering
%and implications for strange form factors,''
Phys.\ Lett.\ B {\bf 509}, 211 (2001)
[arXiv:nucl-ex/0006002].
%%CITATION = NUCL-EX 0006002;%%

%\cite{Batigne:2004ma}
\bibitem{Batigne:2004ma}
G.~Batigne  [G0 Collaboration],
%``G0 experiment status,''
Eur.\ Phys.\ J.\ A {\bf 19}, SUPPL1206 (2004).
%%CITATION = EPHJA,A19,SUPPL1206;%%

%\cite{Maas:2004dh}
\bibitem{Maas:2004dh}
F.~E.~Maas {\it et al.},
%``Evidence for strange quark contributions to the nucleon's form factors at
%Q**2 = 0.108-(GeV/c)**2,''
arXiv:nucl-ex/0412030.
%%CITATION = NUCL-EX 0412030;%%

%\cite{Maas:2004ta}
\bibitem{Maas:2004ta}
F.~E.~Maas {\it et al.}  [A4 Collaboration],
%``Measurement of strange quark contributions to the nucleon's form factors at
%Q**2 = 0.230-(GeV/c)**2,''
Phys.\ Rev.\ Lett.\  {\bf 93}, 022002 (2004)
[arXiv:nucl-ex/0401019].
%%CITATION = NUCL-EX 0401019;%%

%\cite{Baunack:2003ir}
\bibitem{Baunack:2003ir}
S.~Baunack  [A4 Collaboration],
%``Parity-violating electron scattering at MAMI: Strangeness in the nucleon,''
Eur.\ Phys.\ J.\ A {\bf 18}, 159 (2003).
%%CITATION = EPHJA,A18,159;%%

%\cite{Horowitz:1999fk}
\bibitem{Horowitz:1999fk}
C.~J.~Horowitz, S.~J.~Pollock, P.~A.~Souder and R.~Michaels,
%``Parity Violating Measurements of Neutron Densities,''
Phys.\ Rev.\ C {\bf 63}, 025501 (2001)
[arXiv:nucl-th/9912038].
%%CITATION = NUCL-TH 9912038;%%

%\cite{Donnelly:1989qs}
\bibitem{Donnelly:1989qs}
T.~W.~Donnelly, J.~Dubach and I.~Sick,
%``Isospin Dependences In Parity Violating Electron Scattering,''
Nucl.\ Phys.\ A {\bf 503}, 589 (1989).
%%CITATION = NUPHA,A503,589;%%

%\cite{Hammer:1996kx}
\bibitem{Hammer:1996kx}
H.~W.~Hammer, U.~G.~Meissner and D.~Drechsel,
%``Dispersion-theoretical analysis of the nucleon electromagnetic form factors:
%Inclusion of time-like data,''
Phys.\ Lett.\ B {\bf 385}, 343 (1996)
[arXiv:hep-ph/9604294].
%%CITATION = HEP-PH 9604294;%%

%\cite{Ecker:1988te}
\bibitem{Ecker:1988te}
G.~Ecker, J.~Gasser, A.~Pich and E.~de Rafael,
%``The Role Of Resonances In Chiral Perturbation Theory,''
Nucl.\ Phys.\ B {\bf 321}, 311 (1989).
%%CITATION = NUPHA,B321,311;%%

%\cite{bethe1955}
\bibitem{bethe1955}
H.~Bethe and F.~deHoffman, {\em Mesons and Fields} (Row and Peterson, Evanston, 1955), Vol. II.

%\cite{federbush}
\bibitem{federbush}
P.~Federbush, M.~L.~Goldberger, and S.~B.~Treiman, Phys. Rev. {\bf 112}, 642 (1958).

%\cite{Hammer:1999uf}
\bibitem{Hammer:1999uf}
H.~W.~Hammer and M.~J.~Ramsey-Musolf,
%``K anti-K continuum and isoscalar nucleon form factors,''
Phys.\ Rev.\ C {\bf 60}, 045204 (1999)
[Erratum-ibid.\ C {\bf 62}, 049902 (2000)]
[arXiv:hep-ph/9903367].
%%CITATION = HEP-PH 9903367;%%

%\cite{Hammer:1998rz}
\bibitem{Hammer:1998rz}
H.~W.~Hammer and M.~J.~Ramsey-Musolf,
%``Spectral content of isoscalar nucleon form factors,''
Phys.\ Rev.\ C {\bf 60}, 045205 (1999)
[Erratum-ibid.\ C {\bf 62}, 049903 (2000)]
[arXiv:hep-ph/9812261].
%%CITATION = HEP-PH 9812261;%%

%\cite{Ramsey-Musolf:1997qx}
\bibitem{Ramsey-Musolf:1997qx}
M.~J.~Ramsey-Musolf and H.~W.~Hammer,
%``K N scattering and the nucleon strangeness radius,''
Phys.\ Rev.\ Lett.\  {\bf 80}, 2539 (1998)
[arXiv:hep-ph/9705409].
%%CITATION = HEP-PH 9705409;%%

%\cite{Musolf:1996qt}
\bibitem{Musolf:1996qt}
M.~J.~Musolf, H.~W.~Hammer and D.~Drechsel,
%``Nucleon strangeness and unitarity,''
Phys.\ Rev.\ D {\bf 55}, 2741 (1997)
[Erratum-ibid.\ D {\bf 62}, 079901 (2000)]
[arXiv:hep-ph/9610402].
%%CITATION = HEP-PH 9610402;%%

%\cite{Geiger:1996re}
\bibitem{Geiger:1996re}
P.~Geiger and N.~Isgur,
%``Strange hadronic loops of the proton: A quark model calculation,''
Phys.\ Rev.\ D {\bf 55}, 299 (1997)
[arXiv:hep-ph/9610445].
%%CITATION = HEP-PH 9610445;%%

%\cite{Musolf:1996zv}
\bibitem{Musolf:1996zv}
M.~J.~Musolf and H.~Ito,
%``Chiral Symmetry and the Nucleon's Vector Strangeness Form Factors,''
Phys.\ Rev.\ C {\bf 55}, 3066 (1997)
[arXiv:nucl-th/9607021].
%%CITATION = NUCL-TH 9607021;%%

%\cite{Hemmert:1998pi}
\bibitem{Hemmert:1998pi}
T.~R.~Hemmert, U.~G.~Meissner and S.~Steininger,
%``Strange magnetism in the nucleon,''
Phys.\ Lett.\ B {\bf 437}, 184 (1998)
[arXiv:hep-ph/9806226].
%%CITATION = HEP-PH 9806226;%%

%\cite{Hammer:2002ei}
\bibitem{Hammer:2002ei}
H.~W.~Hammer, S.~J.~Puglia, M.~J.~Ramsey-Musolf and S.~L.~Zhu,
%``What do we know about the strange magnetic radius?,''
Phys.\ Lett.\ B {\bf 562}, 208 (2003)
[arXiv:hep-ph/0206301].
%%CITATION = HEP-PH 0206301;%%

%\cite{Dong:1997xr}
\bibitem{Dong:1997xr}
S.~J.~Dong, K.~F.~Liu and A.~G.~Williams,
%``Lattice calculation of the strangeness magnetic moment of the nucleon,''
Phys.\ Rev.\ D {\bf 58}, 074504 (1998)
[arXiv:hep-ph/9712483].
%%CITATION = HEP-PH 9712483;%%

%\cite{Lewis:2002ix}
\bibitem{Lewis:2002ix}
R.~Lewis, W.~Wilcox and R.~M.~Woloshyn,
%``The nucleon's strange electromagnetic and scalar matrix elements,''
Phys.\ Rev.\ D {\bf 67}, 013003 (2003)
[arXiv:hep-ph/0210064].
%%CITATION = HEP-PH 0210064;%%

%\cite{Leinweber:2004tc}
\bibitem{Leinweber:2004tc}
D.~B.~Leinweber {\it et al.},
%``Precise determination of the strangeness magnetic moment of the nucleon,''
arXiv:hep-lat/0406002.
%%CITATION = HEP-LAT 0406002;%%

%\cite{Ji:1995rd}
\bibitem{Ji:1995rd}
X.~D.~Ji and J.~Tang,
%``Locality of the strange sea in the nucleon,''
Phys.\ Lett.\ B {\bf 362}, 182 (1995)
[arXiv:hep-ph/9507465].
%%CITATION = HEP-PH 9507465;%%

%\cite{Silva:2002ej}
\bibitem{Silva:2002ej}
A.~Silva, H.~C.~Kim and K.~Goeke,
%``Strange and singlet form factors of the nucleon: Predictions for G0, A4, and
%HAPPEX-II experiments,''
Eur.\ Phys.\ J.\ A {\bf 22}, 481 (2004)
[arXiv:hep-ph/0210189].
%%CITATION = HEP-PH 0210189;%%

%\cite{Marciano:1983ss}
\bibitem{Marciano:1983ss}
W.~J.~Marciano and A.~Sirlin,
%``On Some General Properties Of The O (Alpha) Corrections To Parity Violation
%In Atoms,''
Phys.\ Rev.\ D {\bf 29}, 75 (1984)
[Erratum-ibid.\ D {\bf 31}, 213 (1985)].
%%CITATION = PHRVA,D29,75;%%

%\cite{Musolf:1990ts}
\bibitem{Musolf:1990ts}
M.~J.~Musolf and B.~R.~Holstein,
%``Electroweak Corrections To Parity Violating Neutral Current Scattering,''
Phys.\ Lett.\ B {\bf 242}, 461 (1990).
%%CITATION = PHLTA,B242,461;%%

%\cite{Musolf:1992xm}
\bibitem{Musolf:1992xm}
M.~J.~Musolf and T.~W.~Donnelly,
%``The Interpretation of parity violating electron scattering experiments,''
Nucl.\ Phys.\ A {\bf 546}, 509 (1992)
[Erratum-ibid.\ A {\bf 550}, 564 (1992)].
%%CITATION = NUPHA,A546,509;%%

%\cite{Musolf:1993tb}
\bibitem{Musolf:1993tb}
M.~J.~Musolf, T.~W.~Donnelly, J.~Dubach, S.~J.~..~Pollock, S.~Kowalski and E.~J.~Beise,
%``Intermediate-energy semileptonic probes of the hadronic neutral current,''
Phys.\ Rept.\  {\bf 239}, 1 (1994).
%%CITATION = PRPLC,239,1;%%

%\cite{Hadjmichael:1991be}
\bibitem{Hadjmichael:1991be}
E.~Hadjmichael, G.~I.~Poulis and T.~W.~Donnelly,
%``Parity violating asymmetry in quasielastic e d scattering,''
Phys.\ Rev.\ C {\bf 45}, 2666 (1992).
%%CITATION = PHRVA,C45,2666;%%

%\cite{Zhu:2000gn}
\bibitem{Zhu:2000gn}
S.~L.~Zhu, S.~J.~Puglia, B.~R.~Holstein and M.~J.~Ramsey-Musolf,
%``The nucleon anapole moment and parity-violating e p scattering,''
Phys.\ Rev.\ D {\bf 62}, 033008 (2000)
[arXiv:hep-ph/0002252].
%%CITATION = HEP-PH 0002252;%%

%\cite{Maekawa:2000qz}
\bibitem{Maekawa:2000qz}
C.~M.~Maekawa and U.~van Kolck,
%``The anapole form factor of the nucleon,''
Phys.\ Lett.\ B {\bf 478}, 73 (2000)
[arXiv:hep-ph/0006161].
%%CITATION = HEP-PH 0006161;%%

%\cite{Maekawa:2000bd}
\bibitem{Maekawa:2000bd}
C.~M.~Maekawa, J.~S.~Veiga and U.~van Kolck,
%``The nucleon anapole form factor in chiral perturbation theory to
%sub-leading order,''
Phys.\ Lett.\ B {\bf 488}, 167 (2000)
[arXiv:hep-ph/0006181].
%%CITATION = HEP-PH 0006181;%%

%\cite{Riska:2000qw}
\bibitem{Riska:2000qw}
D.~O.~Riska,
%``Mesonic anapole form factors of the nucleons,''
Nucl.\ Phys.\ A {\bf 678}, 79 (2000)
[arXiv:hep-ph/0003132].
%%CITATION = HEP-PH 0003132;%%

%\cite{Schiavilla:2002uc}
\bibitem{Schiavilla:2002uc}
R.~Schiavilla, J.~Carlson and M.~Paris,
%``Parity Violating Interactions and Currents in the Deuteron,''
Phys.\ Rev.\ C {\bf 67}, 032501 (2003)
[arXiv:nucl-th/0212038].
%%CITATION = NUCL-TH 0212038;%%

%\cite{Liu:2002bq}
\bibitem{Liu:2002bq}
C.~P.~Liu, G.~Prezeau and M.~J.~Ramsey-Musolf,
%``Hadronic parity violation and inelastic electron deuteron scattering.
%((V)),''
Phys.\ Rev.\ C {\bf 67}, 035501 (2003)
[arXiv:nucl-th/0212041].
%%CITATION = NUCL-TH 0212041;%%

%\cite{Zhu:2001re}
\bibitem{Zhu:2001re}
S.~L.~Zhu, C.~M.~Maekawa, G.~Sacco, B.~R.~Holstein and M.~J.~Ramsey-Musolf,
%``Electroweak radiative corrections to parity-violating electroexcitation  of
%the Delta,''
Phys.\ Rev.\ D {\bf 65}, 033001 (2002)
[arXiv:hep-ph/0107076].
%%CITATION = HEP-PH 0107076;%%

%\cite{Zhu:2002kh}
\bibitem{Zhu:2002kh}
S.~L.~Zhu and M.~J.~Ramsey-Musolf,
%``The off-diagonal Goldberger-Treiman relation and its discrepancy,''
Phys.\ Rev.\ D {\bf 66}, 076008 (2002)
[arXiv:hep-ph/0207304].
%%CITATION = HEP-PH 0207304;%%

%\cite{Siegert:1937yt}
\bibitem{Siegert:1937yt}
A.~J.~F.~Siegert,
%``Note On The Interaction Between Nuclei And Electromagnetic Radiation,''
Phys.\ Rev.\  {\bf 52}, 787 (1937).
%%CITATION = PHRVA,52,787;%%

%\cite{Zhu:2001br}
\bibitem{Zhu:2001br}
S.~L.~Zhu, C.~M.~Maekawa, B.~R.~Holstein and M.~J.~Ramsey-Musolf,
%``Parity violating photoproduction of pi+- on the Delta resonance,''
Phys.\ Rev.\ Lett.\  {\bf 87}, 201802 (2001)
[arXiv:hep-ph/0106216].
%%CITATION = HEP-PH 0106216;%%

%\cite{Arrington:2003qk}
\bibitem{Arrington:2003qk}
J.~Arrington,
%``Implications of the discrepancy between proton form factor  measurements,''
Phys.\ Rev.\ C {\bf 69}, 022201 (2004)
[arXiv:nucl-ex/0309011].
%%CITATION = NUCL-EX 0309011;%%

%\cite{Guichon:2003qm}
\bibitem{Guichon:2003qm}
P.~A.~M.~Guichon and M.~Vanderhaeghen,
%``How to reconcile the Rosenbluth and the polarization transfer method in  the
%measurement of the proton form factors,''
Phys.\ Rev.\ Lett.\  {\bf 91}, 142303 (2003)
[arXiv:hep-ph/0306007].
%%CITATION = HEP-PH 0306007;%%

%\cite{Wells:2000rx}
\bibitem{Wells:2000rx}
S.~P.~Wells {\it et al.}  [SAMPLE collaboration],
%``Measurement of the vector analyzing power in elastic electron proton
%scattering as a probe of double photon exchange amplitudes,''
Phys.\ Rev.\ C {\bf 63}, 064001 (2001)
[arXiv:nucl-ex/0002010].
%%CITATION = NUCL-EX 0002010;%%

%\cite{Maas:2004pd}
\bibitem{Maas:2004pd}
F.~E.~Maas {\it et al.},
% ``Measurement of the transverse beam spin asymmetry in elastic electron proton
%scattering and the inelastic contribution to the imaginary part of the
%two-photon exchange amplitude,''
arXiv:nucl-ex/0410013.
%%CITATION = NUCL-EX 0410013;%%

%\cite{mott}
\bibitem{mott} N.~F.~Mott, Proc. R. Soc. London, Ser. A {\bf 135}, 429 (1932).

%\cite{Diaconescu:2004aa}
\bibitem{Diaconescu:2004aa}
L.~Diaconescu and M.~J.~Ramsey-Musolf,
Phys.\ Rev.\ C {\bf 70}, 054003 (2004)
%``The vector analyzing power in elastic electron proton scattering,''
[arXiv:nucl-th/0405044].
%%CITATION = NUCL-TH 0405044;%%

%\cite{Pasquini:2004pv}
\bibitem{Pasquini:2004pv}
B.~Pasquini and M.~Vanderhaeghen,
%``Resonance estimates for single spin asymmetries in elastic electron nucleon
%scattering,''
Phys.\ Rev.\ C {\bf 70}, 045206 (2004)
[arXiv:hep-ph/0405303].
%%CITATION = HEP-PH 0405303;%%

%\cite{Afanasev:2004pu}
\bibitem{Afanasev:2004pu}
A.~V.~Afanasev and N.~P.~Merenkov,
%``Collinear photon exchange in the beam normal polarization asymmetry of
%elastic electron proton scattering,''
Phys.\ Lett.\ B {\bf 599}, 48 (2004)
[arXiv:hep-ph/0407167].
%%CITATION = HEP-PH 0407167;%%

%\cite{Afanasev:2004hp}
\bibitem{Afanasev:2004hp}
A.~V.~Afanasev and N.~P.~Merenkov,
%``Large logarithms in the beam normal spin asymmetry of elastic electron
%proton scattering,''
Phys.\ Rev.\ D {\bf 70}, 073002 (2004)
[arXiv:hep-ph/0406127].
%%CITATION = HEP-PH 0406127;%%


%\cite{Erler:2003yk}
\bibitem{Erler:2003yk}
J.~Erler, A.~Kurylov and M.~J.~Ramsey-Musolf,
%``The weak charge of the proton and new physics,''
Phys.\ Rev.\ D {\bf 68}, 016006 (2003)
[arXiv:hep-ph/0302149].
%%CITATION = HEP-PH 0302149;%%

%\cite{Kurylov:2003zh}
\bibitem{Kurylov:2003zh}
A.~Kurylov, M.~J.~Ramsey-Musolf and S.~Su,
%``Probing supersymmetry with parity-violating electron scattering,''
Phys.\ Rev.\ D {\bf 68}, 035008 (2003)
[arXiv:hep-ph/0303026].
%%CITATION = HEP-PH 0303026;%%

%\cite{Kurylov:2003xa}
\bibitem{Kurylov:2003xa}
A.~Kurylov, M.~J.~Ramsey-Musolf and S.~Su,
%``Supersymmetric effects in parity-violating deep inelastic electron  nucleus
%scattering,''
Phys.\ Lett.\ B {\bf 582}, 222 (2004)
[arXiv:hep-ph/0307270].
%%CITATION = HEP-PH 0307270;%%

%\cite{Ramsey-Musolf:2000qn}
\bibitem{Ramsey-Musolf:2000qn}
M.~J.~Ramsey-Musolf,
%``Nuclear beta-decay, atomic parity violation, and new physics,''
Phys.\ Rev.\ D {\bf 62}, 056009 (2000)
[arXiv:hep-ph/0004062].
%%CITATION = HEP-PH 0004062;%%

%\cite{Ramsey-Musolf:1999qk}
\bibitem{Ramsey-Musolf:1999qk}
M.~J.~Ramsey-Musolf,
%``Low-energy parity violation and new physics,''
Phys.\ Rev.\ C {\bf 60}, 015501 (1999)
[arXiv:hep-ph/9903264].
%%CITATION = HEP-PH 9903264;%%

\cite{Czarnecki:1995fw,Ferroglia:2003wa,Erler:2004in}

%\cite{Czarnecki:1995fw}
\bibitem{Czarnecki:1995fw}
A.~Czarnecki and W.~J.~Marciano,
%``Electroweak radiative corrections to polarized Moller scattering
%asymmetries,''
Phys.\ Rev.\ D {\bf 53}, 1066 (1996)
[arXiv:hep-ph/9507420].
%%CITATION = HEP-PH 9507420;%%

%\cite{Ferroglia:2003wa}
\bibitem{Ferroglia:2003wa}
A.~Ferroglia, G.~Ossola and A.~Sirlin,
%``The electroweak form factor kappa-hat(q**2) and the running of
%sin**2(theta-hat(W)),''
Eur.\ Phys.\ J.\ C {\bf 34}, 165 (2004)
[arXiv:hep-ph/0307200].
%%CITATION = HEP-PH 0307200;%%

%\cite{Erler:2004in}
\bibitem{Erler:2004in}
J.~Erler and M.~J.~Ramsey-Musolf,
%``The weak mixing angle at low energies,''
arXiv:hep-ph/0409169.
%%CITATION = HEP-PH 0409169;%%

%\cite{Zeller:2001hh}
\bibitem{Zeller:2001hh}
G.~P.~Zeller {\it et al.}  [NuTeV Collaboration],
%``A precise determination of electroweak parameters in neutrino nucleon
%scattering,''
Phys.\ Rev.\ Lett.\  {\bf 88}, 091802 (2002)
[Erratum-ibid.\  {\bf 90}, 239902 (2003)]
[arXiv:hep-ex/0110059].
%%CITATION = HEP-EX 0110059;%%





%Author, \textit{Book title} (Publisher, place year) page numbers
% etc
\end{thebibliography}

\end{document}